# The physical effects of an intra-aggregate structure on soil shrinkage

V.Y. Chertkov*

Division of Environmental, Water, and Agricultural Engineering,
Faculty of Civil and Environmental Engineering, Technion, Haifa 32000, Israel

**Abstract**

Clay and soil containing it have shrinkage curves that are qualitatively different in shape. The objective of this work is to qualitatively show with maximum simplicity, how a clay shrinkage curve turns into a soil shrinkage curve. Because of the crack volume the measured shrinkage curve is not the single-valued feature of a soil. We use a concept of the reference shrinkage curve that is only stipulated by soil shrinkage without cracking, single-valued, and qualitatively similar to an observed shrinkage curve. We also use new concepts of an intra-aggregate soil structure: (i) a rigid superficial layer of aggregates that loses water during shrinkage; and (ii) lacunar pores (micro-cracks) inside an intra-aggregate clay that change in volume during shrinkage. Then, through a series of consecutive steps, illustrating each step by a separate graphic presentation, we move from a clay shrinkage curve to a soil shrinkage curve with predicted qualitative features that coincide with those experimentally observed in numerous soil shrinkage publications. We thereby demonstrate the qualitative physical impact of the intra-aggregate structure on soil shrinkage.

*Corresponding author. Tel.: 972-4829-2601.
*E-mail address:* agvictor@tx.technion.ac.il; vychert@ymail.com (V.Y. Chertkov).

## 1. Introduction

The shrinkage curves of a soil (e.g., Reeve and Hall, 1978; Braudeau et al., 2005) and the clay contributing to it (e.g., Tessier and Pédro, 1984) are qualitatively different (Fig.1) even if the organic matter content is negligible and the clay is the only shrink-swell component of the soil. These differences were recently quantitatively explained for a case of sufficiently high clay content (Chertkov, 2007). The qualitative differences between the shape of clay and aggregated soil shrinkage curves are as follows (Chertkov, 2007; Fig.1):

(a) unlike pure clays (Fig.1a), in the main part of the structural shrinkage area ($W_s<W<W_h$; Fig.1b) the soil shrinkage curve is convex upward;

(b) unlike pure clays (Fig.1a), the slope of the soil shrinkage curve in the structural shrinkage area is essentially less than unity and, in general, varies with water content; the shrinkage curve can even have an inflection point in the area (see Fig.1b, curves 1 and 2) (e.g., Braudeau et al., 2005);

(c) in the basic shrinkage area ($W_n<W<W_s$; Fig.1b) the shrinkage curve slope is constant for a given soil, but unlike the different clays (when the slope is always equal to unity; Fig.1a) for different soils the slope varies between zero and unity; and

(d) unlike pure clays, the maximum swelling point, $W_h$ (or the shrinkage start point) of a soil can be situated at the pseudo saturation line (Fig.1b) (e.g., Braudeau et al., 2005).

In general, the shrinkage curve of an aggregated soil is non-single valued because the volume of (macro) cracks between primary aggregates depends on sampling, sample preparation, sample size, and drying regime (Yule and Ritchie, 1980a, 1980b;

McGarry and Daniels, 1987; Crescimanno and Provenzano, 1999; Braudeau et al., 1999; Cabidoche and Ruy, 2001). However, the qualitative view of the shrinkage curve in Fig.1b is kept at any inter-aggregate cracking. For this reason, to consider the effects of soil structure on shrinkage curve, Chertkov (2007) used the *reference shrinkage curve* that by definition only corresponds to soil clay shrinkage without inter-aggregate cracking and can be predicted in a single valued manner. The negligible volume of inter-aggregate cracks suggests sufficiently small samples and an approximately constant volume of structural (inter-aggregate) pores whose deformation initiates the development of inter-aggregate cracks.

The reference shrinkage curve can be interesting for numerous applications, and first of all for estimating the crack volume contribution to a measured shrinkage curve. In addition, since the reference shrinkage curve differs from the observed shrinkage curve quantitatively, but not qualitatively, it can also be used to reach fundamental understanding of the origin of the abovementioned differences between the shrinkage curves in Fig.1a and Fig.1b.

All the current models of the soil shrinkage curve, at all the differences between them, are based on the approximation of the experimental shrinkage curve data by some a priori taken mathematical expression (different for each different approach) (Groenevelt and Bolt, 1972; Giraldez et al., 1983; Tariq and Durnford, 1993; Olsen and Haugen, 1998; Crescimanno and Provenzano, 1999; Braudeau et al., 1999; Groenevelt and Grant, 2001; Braudeau et al., 2004; Peng and Horn, 2005; Cornelis et al., 2006). Such expressions are not derived from considerations of the soil inter- and intra-aggregate structure, but justified by the fitting of their parameters (from 3 to 11 depending on the approach) to the experimental shrinkage curve data. The a priori accepted mathematical approximation of a shrinkage curve can be useful for applications (e.g., Peng and Horn, 2005; Cornelis et al., 2006) and other aims (e.g., Braudeau et al., 2004). However, explanation and understanding of the characteristic features (see above (a) through (d)) of a soil shrinkage curve and their origin, implies qualitative or quantitative derivation of the shrinkage curve from inter- and intra-aggregate soil structure. Accepting some empirical approximation for the shrinkage curve from the beginning (in the frame of existing models) obviously makes such a derivation meaningless because one can only obtain the approximation that has been a priori given.

Chertkov (2007) relied on the new concepts (and, correspondingly, assumptions) of intra-aggregate soil structure. These concepts were in part inspired by observations from Fiès and Bruand (1998). In general, in addition to the structural pores (associated with inter-aggregate space) and clay pores (associated with the space between clay particles), the soils also contain the micro-cracks or *lacunar pores* inside a clay of primary aggregates (Fiès and Bruand, 1998). These researches observed lacunar pores (Fig.2) in clay matrices entering the oven-dried artificial clay-silt-sand mixtures. At sufficiently small clay content, $c$ the lacunar pores are three-dimensional, have an appreciable volume (up to that of inter-grain space), and form a network. By increasing the clay content the lacunar pore network gradually disappears, and only isolated (hidden) lacunar pores remain. At sufficiently high clay content the lacunar pores have a crack-like shape and negligible volume.

The consideration of the reference shrinkage curve (Chertkov, 2007) was based on three assumptions.

*Assumption 1*. At a given clay type and porosity $p$ of silt and sand grains coming into contact, there is such a critical clay content, $c^*$ that at a soil clay content $c>c^*$ lacunar pores inside clay and grain contacts are lacking (Fig.2a), but at a soil clay content $c<c^*$ lacunar pores exist and grain contacts can exist, at least at sufficiently small water contents (Fig.2b). If $c<c^*$ and the lacunar pores (Fig.2b) exist they can reach the sizes of inter-grain spaces and are usually large compared to clay pores and emptied (if water-filled) before them. Observations (Fiès and Bruand, 1998) show that $c^*$ can be between ~0.3 and ~0.8, but a typical value $c^* \cong 0.4$ (Revil and Cathles, 1999). The critical value $c^*$ to be defined by the above assumption is expressed through porosity $p$, the relative volume of the clay solids, $v_s$, and the relative oven-dried clay volume, $v_z$ (the $v_s$ and $v_z$ values are connected with clay type, see Chertkov, 2000, 2003) and is in agreement with data from Fiès and Bruand (1998).

*Assumption 2.* The shrinkage of an intraaggregate clay, aggregates, and soil, as a whole, starts simultaneously at total soil water content, $W_h$ corresponding to the point of maximum soil swelling or point of shrinkage start (Fig.1b).

*Assumption 3.* Water in the clay pores of a thin rigid superficial layer of aggregates (Fig.2) with modified pore structure (interface layer) and the layer volume, determine a soil reference shrinkage curve in the structural shrinkage area.

The quantitative consideration (Chertkov, 2007) that only relates to case $c>c^*$ (where lacunar pores are lacking) is quite complex. The quantitative consideration for case $c<c^*$ requires an additional complication. At the same time, based on the above concepts and assumptions, one can propose the relatively simple *qualitative* explanation of the above shape features of the soil reference shrinkage curve originating from those of the clay shrinkage curve. Thus, unlike the quantitative consideration from Chertkov (2007) (for soils with sufficiently high clay content only) that was conducted using a number of inevitable complications, and accounting for the interdisciplinary character of the journal, the objective of this work is to show that a *simple qualitative physical* explanation and understanding of the transformation of a clay shrinkage curve (Fig.1a) to the reference shrinkage curve of a soil (Fig.1b) with any clay content and with the above characteristic features from (a) to (d), can be reached without any calculations and only using rather simple means. In addition, our aim is to illustrate the approach with maximum possible simplicity and clearness, step by step, using a separate graphic presentation for each step (unlike Chertkov (2007) where the steps are shown together, in one complex figure).

Correspondingly, the data that we are going to use for comparison with theoretical results, relate to the *qualitative* aspects of the shrinkage curve shape. The qualitative aspects were formulated above as characteristic features from (a) to (d) (Chertkov, 2007) and visually shown in Fig.1. As for the experimental observations of these qualitative aspects, the reader can turn to numerous soil shrinkage publications, major of which have already been mentioned above. Notation is summarized in Section 9.

**2. The shrinkage curve of a pure clay** ($V(\overline{w})$)

We start from a disaggregated clay. The shrinkage curve of a disaggregated clay can be written as

$$V(\overline{w}) = 1/\rho_s + V_{cp}(\overline{w}) \qquad (1)$$

where $V$ is the specific volume of the clay; $\overline{w}$ - the gravimetric water content of the clay; $\rho_s$ - the density of solids (clay particles); and $V_{cp}$ - the specific volume of the clay (matrix) pores. In the frames of Chertkov's (2000, 2003) model Eq.(1) gives (see Fig.1a, solid line shows a qualitative view of $V(\overline{w})$)

$$V(\overline{w}) = \begin{cases} V_z \equiv 1/\rho_s + V_{cp}(\overline{w}_z), & 0 \le \overline{w} \le \overline{w}_z \\ V_z + a(\overline{w} - \overline{w}_z)^2, & \overline{w}_z \le \overline{w} \le \overline{w}_n \\ 1/\rho_s + \overline{w}/\rho_w, & \overline{w}_n \le \overline{w} \le \overline{w}_M \end{cases} \quad (2)$$

where $\overline{w}_z$, $\overline{w}_n$, and $\overline{w}_M$ are the clay shrinkage limit, air-entry point, and the liquid limit, respectively (at natural swelling only the maximum swelling point, $\overline{w}_h$ ($<\overline{w}_M$) is reached); $\rho_w$ – the water density; and $a$ – a coefficient depending on $V_z$, $w_M$, and $\rho_s$ (Chertkov, 2000, 2003).

**3. The shrinkage curve of an intra-aggregate matrix** ($U(w)$)

The aim of this section is to show the link between the shrinkage curve of an intra-aggregate matrix (Fig.2) and the above shrinkage curve of a clay that contributes to it.

*3.1. Lacunar pore volume* ($U_{lp}(w)$) *and lacunar factor* ($k$)

The specific volume $U(w)$ of the intra-aggregate matrix (Fig.2a and 2b) where $w$ is the gravimetric water content of the matrix, includes the constant specific volume $1/\rho_s$ of solids as well as the specific volumes of lacunar pores $U_{lp}$ and clay pores $U_{cp}$ as (note that here we rely on the assumptions from Section 1)

$$U(w)=1/\rho_s+U_{lp}(w)+U_{cp}(w), \quad 0 \le w \le w_h \quad (3)$$

where $w_h$ is the maximum swelling point of the intra-aggregate matrix at natural swelling ($w_h<w_M$ where $w_M$ is the liquid limit of the intra-aggregate matrix). In Eqs.(2) and (3) for the sake of simplicity we do not take into account a possible small difference between the density of silt-sand grains and clay solids, and correspondingly, the difference between the clay content by weight and the volume fraction of clay solids. The corresponding modification is possible but introduces a superfluous complication.

The small variation $dU_{cp}$ of the clay pore volume at shrinkage ($dU_{cp}<0$), corresponding to a water loss $dw<0$, initiates both the variation $dU$ of the intra-aggregate matrix volume ($dU<0$) and the variation $dU_{lp}$ of the lacunar pore volume ($dU_{lp}>0$ since the shrinkage of intra-aggregate clay leads to the lacunar pore volume increase). According to Eq.(3) the obvious volume balance takes place as

$$dU_{cp}=dU - dU_{lp}, \quad 0 \le w \le w_h. \quad (4)$$

Based on the balance relation one can present $dU_{lp}$ and $dU$ as

$$dU_{lp}=-k\, dU_{cp}, \quad 0 \le w \le w_h \quad (5)$$

$$dU=(1-k)\, dU_{cp}, \quad 0 \le w \le w_h. \quad (6)$$

In general, in Eqs.(5) and (6) the lacunar factor, $k$ is a function of clay and water content, $k(c, w)$ with values varying in the range $0 \le k \le 1$. At a high clay content ($c>c^*$) the lacunar pore volume is negligible (Fig.2a) and according to Eq.(5) $k=0$ in the whole range $0 \le w \le w_h$. At $c<c^*$ lacunar pores exist and according to Eqs.(5) and (6) $0<k \le 1$. We assume that at $c<c^*$ as in the case $c>c^*$ when $k=0=$constant as a function of $w$, $k$ does not also depend on water content, but can vary with clay content in the above range $0<k\leq1$. We will see that experimental observations justify this assumption. Then, integrating Eq.(5) at any clay content over water content from the initial $w_h$ value to a current $w$ value, and taking into account the initial conditions, $U_{lp}(w_h)\equiv U_{lph}$ and $U_{cp}(w_h)=w_h/\rho_w$ as well as the fact that $k$ does not depend on water content, we find for any given clay content a linear relation between the lacunar pore volume ($U_{lp}(w)$) and the clay pore volume ($U_{cp}(w)$) at shrinkage to be

$$U_{lp}(w)=U_{lph}+k(w_h/\rho_w-U_{cp}(w)), \qquad 0\leq w\leq w_h . \tag{7}$$

*3.2. The shrinkage curve of an intra-aggregate matrix ($U(w)$) vs. the shrinkage curve of a clay ($V(\overline{w})$) (Fig.3)*

Although lacunar pores change in volume (Eq.(7)) during shrinkage they are emptied (if water filled) before the shrinkage starts (because of their size and capillarity). That is, at any water content $0\leq w\leq w_h$ of the intra-aggregate matrix, water is only contained in clay pores (Fig.2). It follows a simple relation between the gravimetric water content ($\overline{w}$) of a pure clay (per unit mass of oven-dried clay) and that ($w$) of the intra-aggregate matrix (Fig.2) (per unit mass of oven-dried matrix) containing the clay as

$$w=\overline{w}\,c , \qquad 0\leq \overline{w}\leq \overline{w}_h, \quad 0\leq w\leq w_h \tag{8}$$

(where $c$ is the clay content of the matrix and soil) and a similar relation

$$U_{cp}(w)=V_{cp}(\overline{w})\,c , \qquad 0\leq \overline{w}\leq \overline{w}_h, \quad 0\leq w\leq w_h \tag{9}$$

between the specific volume $V_{cp}$ of clay pores (per unit mass of oven-dried clay) and specific volume $U_{cp}$ of clay pores of the intra-aggregate matrix (Fig.2) (per unit mass of oven-dried intra-aggregate matrix) containing the clay. In addition, from Eq.(1)

$$V_{cp}(\overline{w})=V(\overline{w})-1/\rho_s . \tag{10}$$

Thus, Eqs.(9), (10), and (8) determine $U_{cp}(w)$ through $V(w/c)$. Similarly, Eqs.(7), (9), (10), and (8) determine $U_{lp}(w)$ through $V(w/c)$. Finally, Eqs.(3), (7), (9), (10), and (8) determine the specific volume of the intra-aggregate matrix, $U(w)$ through the specific volume $V(w/c)$ of the clay contributing to the matrix (Fig.2) as

$$U(w)=1/\rho_s+U_{lph}+k\,w_h/\rho_w+(1-k)c(V(w/c)-1/\rho_s) , \qquad 0\leq w\leq w_h . \tag{11}$$

Eq.(11) relates to any clay content. At $c<c^*$ in Eq.(11) $U_{lph}>0$ and $0<k=\text{const}\leq1$. At $c>c^*$ in Eq.(11) $U_{lph}=0$ and $k=0$. Thus, there is the linear relation between $U(w)$ and $V(w/c)$. Replacing $V(w/c)$ in Eq.(11) from Eq.(2) (Chertkov, 2000) one can obtain the following presentation of $U(w)$

$$U(w)=\begin{cases} U_z, & 0\leq w\leq w_z \\ U_z+b(w-w_z)^2, & w_z\leq w\leq w_n \\ 1/\rho_s+U_{lph}+k\,w_h/\rho_w+(1-k)w/\rho_w, & w_n\leq w\leq w_h \end{cases} \tag{12}$$

$U_z$ and $b$ are expressed through clay parameters (Eq.(2)) and the intra-aggregate matrix parameters (Eq.(11)). Here we are only interested in the qualitative view of $U(w)$ in comparison with the $V(\overline{w})$ dependence in Fig.1a. Fig.3 shows such a view of $U(w)$ at $0<k<1$ (i.e., for $c<c^*$ and $U_{lph}>0$). The $w$ value along the water content axis in Fig.3 is connected with $\overline{w}$ in Fig.1a by Eq.(8). In particular, $w_z=\overline{w}_z c$, $w_n=\overline{w}_n c$, and $w_h=\overline{w}_h c$.

At first glance the presentations of Eqs.(2) and (12) as well as dependences $V(\overline{w})$ in Fig.1a and $U(w)$ in Fig.3 are similar. Indeed, in both cases Fig.1a and Fig.3 there are two linear sections and a "squared" section. However, one can also see two very essential qualitative differences between $U(w)$ (Eq.(12)) and $V(\overline{w})$ (Eq.(2)). The former difference relates to the slope of the $U(w)$ curve in the basic shrinkage area ($w_n \leq w \leq w_h$) (Fig.3) compared with the slope of the $V(\overline{w})$ curve in its basic shrinkage area ($\overline{w}_n \leq \overline{w} \leq \overline{w}_h$) (Fig.1a). The slope $dV/d\overline{w}=1/\rho_w$ at $\overline{w}_n \leq \overline{w} \leq \overline{w}_h$ (Eq.(2)) for any clay (Fig.1a). Unlike that, according to Eq.(12) the slope

$$dU/dw = (1-k)/\rho_w, \qquad w_n \leq w \leq w_h \tag{13}$$

and depends on the $k$ value (see Fig.3). In turn, the $k$ factor depends on the clay content and characteristics of clay and silt-sand grains contributing to the soil (see Section 3.1). Only at $c>c^*$ $dU/dw=1/\rho_w$ in Fig.3 in the total range $w_n \leq w \leq w_h$ similar to clay (Fig.1a) because $k=0$. At $c<c^*$ (Fig.3) the slope is constant and less than unity (($1-k)/\rho_w$) in the total range $w_n \leq w \leq w_h$. Equation (13) can be directly obtained from Eq.(6) because at $w_n \leq w \leq w_h$ $dU_{cp}=dw/\rho_w$. The above analysis is necessary to understand the relations between $V(\overline{w})$ and $U(w)$.

The second qualitative difference relates to the position of the saturation line of the $U(w)$ curve (Fig.3) unlike the position of the saturation line of the $V(\overline{w})$ curve (Fig.1a). In the range $\overline{w}_n \leq \overline{w} \leq \overline{w}_h$ the shrinkage curve $V(\overline{w})$ of a disaggregated clay always coincides with its saturation line (see Eq.(2) and Fig.1a). Unlike that, the point $U_h=U(w_h)$ of the shrinkage curve $U(w)$ of the intra-aggregate matrix (Fig.3) is only situated on the true saturation line, $1/\rho_s+w/\rho_w$ if $k=0$ and $U_{lph}=0$ (see Eq.(12) at $w_n \leq w \leq w_h$). This is realized for clay contents $c>c^*$ and corresponds to the absence of lacunar pores. If $k$=const>0 (Fig.3) lacunar pores exist, their volume remains empty, and the true saturation line, $1/\rho_s+w/\rho_w$ is not reached at swelling. In this case ($c<c^*$) the point $U_h=U(w_h)$ of the shrinkage curve $U(w)$ (Fig.3) is situated on a 1:1 line (with unit slope) that is a pseudo saturation line.

It is worth emphasizing that both the above qualitative differences between $U(w)$ and $V(\overline{w})$ are direct consequences of lacunar pores existing in the intra-aggregate matrix (Fig.2b) and the increase in lacunar pore volume ($U_{lp}(w)$) (Eqs.(7)-(10)) during shrinkage. At high clay content, $c>c^*$ when there are no lacunar pores (Fig.2a), these qualitative features of $U(w)$ compared with $V(\overline{w})$ are absent; that is, the dashed and dash-dot lines in Fig.3 coincide with each other and with $U(w)$ at $w_n \leq w \leq w_h$.

Note that at high clay content ($c>c_*$) the shrinkage curve of the intra-aggregate matrix, $U(w)$ (Fig.3) can be directly observed as the shrinkage curve of a clay paste with silt-sand admixture (e.g., Bruand and Prost, 1987). However, at sufficiently small clay content (at $c<c^*$) the shrinkage curve $U(w)$ (Fig.3), it seems, cannot be directly observed because of the quick formation of a superficial aggregate layer that is present in any sample along with an intra-aggregate matrix (Fig.2b).

Finally, note that in the following consideration the shrinkage curve of the intra-aggregate matrix, $U(w)$ plays the part of an important auxiliary curve.

**4. The shrinkage curve of an aggregate system ($U_a(W)$)**

The shrinkage curve of an aggregate system is a correspondence between the total gravimetric water content ($W$) and specific volume ($U_a$) of aggregates. Both consist of two contributions (Chertkov, 2007). The aim of this section is to show, step by step, the transition from the $U(w)$ auxiliary curve to the shrinkage curve $U_a(W)$.

*4.1. Contributions to the total water content ($W$) and volume ($U_a$) of aggregates*

According to Fig.2 and assumptions from Section 1 the total water content of aggregates (and soil) ($W$) includes contributions of the intra-aggregate matrix ($w'$) and interface layer ($\omega$) as

$$W = w' + \omega . \tag{14}$$

Structural and lacunar pores are empty at shrinkage because of their size and capillarity. Thus, the $w'$ and $\omega$ contributions are interconnected because the interface clay pores giving the $\omega$ contribution and pores of intra-aggregate clay giving the $w'$ contribution (Fig.2) are at the same varying suction. We will consider $w'$ to be an independent variable that varies in the range $0 \leq w' \leq w_h'$ where $w_h'$ corresponds to the state of the maximum aggregate and soil swelling.

According to Fig.2 and assumptions from Section 1 the specific volume of aggregates ($U_a$) also includes contributions of the intra-aggregate matrix ($U'$) and rigid interface layer ($U_i$). $U'$ and $U_a$ can also be considered as functions of the independent parameter $w'$. Thus,

$$U_a(w') = U'(w') + U_i , \qquad 0 \leq w' \leq w_h' . \tag{15}$$

*4.2. Contribution of the intra-aggregate matrix ($U'(w')$) to the aggregate volume ($U_a$) vs. the shrinkage curve of the intra-aggregate matrix ($U(w)$) (Fig.4)*

The $U'(w')$ dependence in itself (Fig.4) is not a real shrinkage curve because it only gives one contribution (of the intra-aggregate matrix; Fig.2) to the specific volume, $U_a$ (another one is $U_i$, Eq.(15)). This is why the $U'(w')$ curve is situated below the saturation line in Fig.4. However, $U'(w')$ is directly and simply connected with the auxiliary shrinkage curve of the intra-aggregate matrix, $U(w)$ (Fig.3; Section 3). $U'(w')$ and $U(w)$ only differ by normalization. $U'$ and $w'$ are the specific volume and water content contributions of the intra-aggregate matrix per unit mass of the oven-dried soil as a whole, including the solid mass of the interface layer (see Fig.2). $U$ and $w$, however, are the specific volume and water content of the same intra-aggregate matrix per unit mass of the oven-dried matrix itself, i.e., without the solid mass of the interface layer (see Fig.2). It follows that the water content values along the $w$ and $w'$ axes in Fig.4 are connected as

$$w' = w/K , \tag{16}$$

and corresponding $U$ and $U'$ values as

$$U' = U/K \tag{17}$$

where $K>1$ is the ratio of the aggregate solid mass to the solid mass of the intra-aggregate matrix (i.e., the solid mass of aggregates without the interface layer, see Fig.2). Hence, the relation between $U'(w')$ and $U(w)$ is

$$U'(w')=U(w'K)/K, \qquad 0\leq w'\leq w_h' \ . \tag{18}$$

Thus, the $U'(w')$ curve (Fig.4) is expressed through the auxiliary shrinkage curve $U(w)$ (Section 3) and $K$ ratio. The latter is connected with the $U_i$ contribution (see Section 4.3).

According to Eqs.(16) and (18) $dU'/dw'=dU/dw$ (Fig.4; $0\leq w'\leq w_h'$; $0\leq w\leq w_h=Kw_h'$). In particular, the basic shrinkage area, $w_n\leq w\leq w_h$ of $U(w)$ (Fig.4) where $dU/dw=(1-k)/\rho_w$ (Eq.(13)) corresponds to the basic shrinkage area, $w_n'=w_n/K\leq w'\leq w_h'$ of $U'(w')$ (Fig.4) where $U'(w')$ has the similar slope, $(1-k)/\rho_w$. For the sake of simplicity Fig.4 only shows the $U(w)\rightarrow U'(w')$ transition for the case $c<c^*$ when $0<k\leq 1$ (Fig.3). At $k=0$ ($c>c^*$) the slope of $U(w)$ and $U'(w')$ in corresponding basic shrinkage areas coincide with the slope of the saturation line (the dashed line in Fig.4 that at $k=0$ coincides with dash-dot line).

*4.3. The shrinkage curve of aggregate volume in coordinates $U_a$ and $w'$ vs. the contribution of intra-aggregate matrix ($U'(w')$) (Fig.5)*

According to Eq.(15) the shrinkage curve $U_a(w')$ only differs from the $U'(w')$ curve (Section 4.2, Fig.4) by the constant shear $U_i$ along the axis of the specific volume (Fig.5). This means that all that was said in Section 4.2 with respect to the $dU'/dw'$ slope in the basic shrinkage area $w_n'\leq w'\leq w_h'$ (Fig.4) and the slope value connection with clay content, also relates to the $dU_a/dw'$ slope in the same area (Fig.5).

The $U_a(w')$ curve (Fig.5) has an uncustomary view because (unlike $U'(w')$) it was presented using an unusual (for $U_a$ volume) $w'$ coordinate. The presentation in usual coordinates, $U_a(W)$ is determined by a $W(w')$ dependence that changes the scale along the water content axis compared to $w'$ (see Section 4.4).

In connection with the mutual situation of the $U(w)$ curve (Fig.3 and Fig.4) and $U_a(w')$ curve (Fig.5), it is important to emphasize that the $U_h=U(w_h)$ (Fig.3-5) and $U_{ah}=U_a(w_h')$ (Fig.5) values coincide as

$$U_{ah}=U_h \tag{19}$$

because the water contents and specific volumes of the two aggregate parts - the rigid interface layer and intra-aggregate matrix (Fig.2) - coincide at maximum soil swelling (the shrinkage start point). Replacing in relation, $U_i=U_{ah}-U_h'$ (see Eq.(15) at $w'=w_h'$) $U_{ah}$ from Eq.(19) and $U_h'=U_h/K$ (from Eq.(17) at $w'=w_h'$) we obtain the important relation between the specific volume of the interface layer, $U_i$ (Fig.2 and Fig.5), the specific volume of the intra-aggregate matrix ($U_h$) (or specific volume of aggregates ($U_{ah}$) as a whole) at maximum swelling, and the $K$ ratio as

$$U_i=U_h(1-1/K) \ . \tag{20}$$

This relation reflects the mutual arrangement of curves $U(w)$, $U'(w')$, and $U_a(w')$ in Fig.3-5.

For the sake of simplicity Fig.5 only shows the $U'(w') \to U_a(w')$ transition for the case $c<c^*$ when $0<k\leq 1$ and $U(w)$ corresponds to Fig.3. At $k=0$ ($c>c^*$) the slope of $U'(w')$ and $U_a(w')$ in Fig.5 in the $w_n'\leq w'\leq w_h'$ area coincide with the slope of the dashed and dash-dot lines that also coincide with each other.

*4.4. The shrinkage curve of an aggregate system in coordinates $U_a$ and $W$ vs. the same shrinkage curve in coordinates $U_a$ and $w'$ (Fig.7)*

We consider $w'$ to be an independent variable (Section 4.1). Then the contribution of the interface layer to the total water content $W$ is $\omega=\omega(w')$ (see Eq.(14)). This dependence can be found quantitatively for the two possible types of modified pore-size distribution in the interface clay (Chertkov, 2007) from the pore-size distribution of the clay (Chertkov, 2000, 2005) and from the dependence of the maximum size of water-filled clay pores on water content $w'$ (Chertkov, 2004). However, the qualitative view of the $\omega(w')$ dependence (Fig.6) is sufficient for the aims of this work. This qualitative view flows out of simple considerations. $\omega(w_h')=\omega_h$ (Fig.6) gives the maximum water content of the rigid interface layer when shrinkage starts. The interface clay pores are totally emptied during shrinkage at some $w_s'$ (Fig.6) where $w_n'<w_s'<w_h'$ ($w_n'$ is the air-entry point of the intra-aggregate matrix; see Fig.5), i.e., $\omega(w')=0$ at $w'\leq w_s'$ (Fig.6). The $w_s'$ point is the final one of structural shrinkage and the initial one of basic shrinkage (Fig.7). It is obvious that the $\omega(w')$ dependence should be smooth at $w'=w_s'$ ($\left. d\omega/dw' \right|_{w'=w_s'} = 0$; Fig.6). Then, the two curves in Fig.6 show two of the simplest, geometrically possible variants of such $\omega(w')$ dependence with (curve 1) and without (curve 2) inflection point.

The total water content $W$ as a function of $w'$ is $W(w')=w'+\omega(w')$ (Eq.(14)) where $\omega(w')$ is from Fig.6. The $W(w')$ dependence changes the scale along the water content axis compared to $w'$ (Fig.7). At $0\leq w'\leq w_s'$ $W=w'$ (Fig.7). This means that $U_a(w')$ at $0\leq w'\leq w_s'$ in Fig.5 and Fig.7 quantitatively coincides with $U_a(W)$ at $0\leq W\leq W_s$ (i.e., in particular, in the basic shrinkage area, $w_n'\leq w'\leq w_s'$) in Fig.7. Only in the $w_s'\leq w'\leq w_h'$ and $W_s\leq W\leq W_h$ ranges of the structural shrinkage do the presentations $U_a(w')$ and $U_a(W)$ (Fig.7) differ. The two variants of $U_a(W)$ in Fig.7 correspond to the two variants of $\omega(w')$ in Fig.6. Thus, the model predicts two possible qualitatively different types of reference shrinkage curve in the structural shrinkage area. Note that both the convex upward shape of $U_a(W)$ in the structural shrinkage area ($W>W_s$) and two types of $U_a(W)$ in the structural shrinkage area (Fig.7) originate from the existence of the rigid interface layer ($U_i$ addition to $U_a$, Eq.(15)) and its dewatering (Fig.6) depending on its pore-size distribution.

As noted above, at $W\leq W_s$ and $w'\leq w_s'$, $W=w'$ (Fig.7) and $U_a(W)=U_a(w')$. Hence, the slope of the observed reference shrinkage curve, $dU_a/dW$ in the observed basic shrinkage area, $W_n\leq W\leq W_s$ ($w_n'\leq w'\leq w_s'$, see Fig.7) is also given by $(1-k)/\rho_w$ as $dU'/dw'$ and $dU_a/dw'$ (see the end of Section 4.2 and beginning of Section 4.3). This means that all that was said in Section 4.2 with respect to the $dU'/dw'$ slope in the basic shrinkage area $w_n'\leq w'\leq w_h'$ (Fig.4) and the slope value connection with clay content, also relate to the $dU_a/dW$ slope in the same area, $W_n\leq W\leq W_s$ (Fig.7).

In connection with the mutual situation of the $U(w)$ curve (Fig.3 and Fig.4), the $U_a(w')$ curve (Fig.5), and the $U_a(W)$ curve (Fig.7), it is important to emphasize that $w_h$ (Fig.3-5) and $W_h$ (Fig.7) coincide as

$W_h=w_h$ (21)

because the water contents and specific volumes of the two aggregate parts - the rigid interface layer and intra-aggregate matrix (Fig.2) - coincide at maximum soil swelling (the shrinkage start point). Accounting for $w_h=Kw_h'$ (Eq.(16)) and $W_h-w_h'=\omega_h$ (Eq.(14)) Eq.(21) leads to a relation that is equivalent to Eq.(20).

Finally, in the case of the sufficient connectivity of lacunar pores the $U_a(W)$ shrinkage curve can reach the true saturation line and have a horizontal section as a result of their filling at water contents $W_h \leq W \leq W_h^*$ (Fig.7).

For the sake of simplicity Fig.7 only shows the $U_a(w') \rightarrow U_a(W)$ transition for the case $c<c^*$ when $0<k\leq1$, and the $U_a(W)$ curve starts at a pseudo saturation line (at $W=W_h$).

## 5. The shrinkage curve of a soil ($Y(W)$) vs. the aggregate shrinkage curve ($U_a(W)$) (Fig.8)

The specific volume of soil, $Y(W)$ only differs from the specific volume of an aggregate system, $U_a(W)$ by the specific volume $U_s$ of the inter-aggregate (structural) pores as

$$Y(W)=U_a(W)+U_s . \tag{22}$$

In the case of the reference shrinkage curve (by its definition) $U_s$ does not depend on water content $W$ (see Section 1). Thus, the soil reference shrinkage curve, $Y(W)$ is totally similar in shape to the aggregate shrinkage curve, $U_a(W)$ (Fig.8) with the addition of a horizontal section in the range $W_h \leq W \leq W_m$. This section corresponds to the emptying of the structural pores (if water filled), and relations $U_s=Y_z-U_{az}=(W_m-W_h)/\rho_w$ (see Fig.8) take place.

Depending on lacunar pore connectivity the $Y(W)$ curve can start on a pseudo saturation line (at $W=W_m$ in Fig.8) or reach the true saturation line and have an additional horizontal section in the range $W_m \leq W \leq W_m^*$ (Fig.8) as a result of lacunar pore filling.

In the force of the similarity between $Y(W)$ and $U_a(W)$ (Fig.8; Eq.(22)) it is obvious that in the basic shrinkage area ($W_n \leq W \leq W_s$) the slope $dY/dW=dU_a/dW=(1-k)/\rho_w$ and also depends on the range of clay content: $c<c^*$, ($0<k\leq1$; the lacunar pore volume is increasing with drying) or $c>c^*$ ($k=0$; lacking lacunar pores). It also follows from the same similarity that both the convex upward shape of $Y(W)$ in the structural shrinkage area ($W>W_s$) and the two types of $Y(W)$ in the structural shrinkage area (Fig.8 only shows type 2 of $Y(W)$) originate from the existence of the rigid interface layer of aggregates and its dewatering (Fig.6) depending on its pore-size distribution.

Worthy of special note is that the experimentally observed constancy of the shrinkage curve slope of a soil in the basic shrinkage area (feature (c) in Section 1), in combination with the theoretically predicted expression for the slope ($dY/dW=(1-k)/\rho_w$), do justify the above important assumption (Section 3.1) according to which the lacunar factor, $k$ does not depend on water content.

Finally, note that the $U_s \cong 0$ value reflects a broad spectrum of practically important situations when the specific volume of structural pores, $U_s$ is rather less than the specific volume of interface clay pores, $U_i \Pi$ ($\Pi$ is the rigid porosity of the interface clay). In this case $Y(W) \cong U_a(W)$ (Fig.7 and 8). If $U_s << U_i \Pi$ the structural water content (before shrinkage starts) is rather less than the interface water content.

For the sake of simplicity Fig.8 only shows the transition from $U_a(W)$ to $Y(W)$ in the case that was indicated in Fig.1b when $c<c^*$, $0<k\leq1$, and the $Y(W)$ curve starts at the pseudo saturation line.

**6. Data to be used**

It is worth reiterating that the data that we use for comparison with theoretical results, relate to the *qualitative* aspects of the shrinkage curve shape (characteristic features from (a) to (d) from Section 1). Corresponding data are available from numerous soil shrinkage publications, for instance, from Reeve and Hall (1978), Baer and Anderson (1997), and Braudeau et al. (2005). The *qualitative* aspects of the shrinkage curve shape from the available publications are presented in Fig.1 in a generalized form.

**7. Results and discussion**

We assumed two new features of the intra-aggregate structure - the rigid surface (interface) aggregate layer with varying water content and intra-aggregate lacunar pores with varying volume (Fig.2). Then, we visually showed (Figs.3-8) that the features lead to a number of qualitative peculiarities of the shape of the reference shrinkage curve of an aggregated soil. These qualitative physical effects of the new intra-aggregate structure concepts are as follows:

(i) The existence of the rigid superficial aggregate layer (Fig.2) (interface layer) leads to the convex upward shape of the reference shrinkage curve in the structural shrinkage area (see the sequence of Figs.3-8).

(ii) Dewatering of the rigid interface layer of aggregates (Fig.6) depending on its pore-size distribution leads to the specific course (with varying slope <1) and two possible types of reference shrinkage curve in the structural shrinkage area (see curves 1 and 2 in Fig.7).

(iii) The existence and increase of the lacunar pore volume inside the intra-aggregate matrix (Fig.2b) with a decrease in water content and depending on clay content, lead to the reference shrinkage curve slope being equal to or less than unity in the basic shrinkage area (see the sequence of Figs.3-8).

(iv) The existence and increase of the lacunar pore volume inside the intra-aggregate matrix (Fig.2b) with a decrease in water content and depending on clay content, leads to a possible shear of the true saturation line relative to the pseudo saturation line (see the sequence of Figs.3-8).

One can now compare the results of predicting the qualitative peculiarities of the shrinkage curve shape (from (i) to (iv)) with results of numerous observations (see Section 6) that were formulated in Section 1 as points (a)-(b) (Chertkov, 2007) and as visually shown in Fig.1. The comparison reveals the total coincidence between the predicted ((i)-(iv)) and observed ((a)-(d)) peculiarities of the shape of the shrinkage curve.

Thus, a major result of this work is the qualitative explanation of the origin of the observed qualitative differences between a clay shrinkage curve and the shrinkage curve of the soil containing it, based on the new assumed features of an intra-aggregate structure: the existence of lacunar pores and the increase in their volume during shrinkage as well as the existence of the rigid superficial layer of aggregates and its dewatering during shrinkage.

The qualitative explanation of the $V(\overline{w})$ to $Y(W)$ transition is important for the physical understanding of the interconnections between an intra-aggregate structure and a soil shrinkage curve, even though one can quickly measure the shrinkage curve and even though there are good fitting approximations for the shrinkage curve. In general, observed shrinkage curves can quantitatively differ from the reference shrinkage curve due to crack existence (for sufficiently large samples), but qualitatively they have the same characteristic features that are explained above.

## 8. Conclusion

This work shows how a clay shrinkage curve turns into a soil shrinkage curve. Assumptions about an increase in lacunar pore volume and dewatering of a rigid superficial aggregate layer at soil shrinkage enable one to qualitatively explain the origin of the observed peculiarities of the shrinkage curve shape of aggregated soils. Conversely, these observed peculiarities suggest the indicated new features of an intra-aggregate structure. The physical understanding of the transformation of the clay shrinkage curve to the soil shrinkage curve and of the new features of the intra-aggregate structure underlying this transformation, are essential for numerous applications; first of all, for the physical modeling of soil shrinkage, water retention, hydraulic conductivity, water flow, and solute transport. In addition, the concepts and results of the model can also be useful for the physical explanation of crack contribution into a measured shrinkage curve, the hysteretic shrinkage-swelling phenomenon, the effects of compaction and consolidation on soil shrinkage-swelling in the field, the effects of the non-total closing shrinkage cracks at soil wetting in field conditions, and soil crusting.

## 9. Notation

$a$      coefficient in Eq.(2) ($cm^3 g^{-1}$)
$b$      coefficient in Eq.(12) ($cm^3 g^{-1}$)
$c$      weight fraction of clay solids of the total solids (clay-silt-sand) (dimensionless)
$c^*$      first critical clay content, at $c>c^*$ lacunar pores are absent at any water content (dimensionless)
$K$      ratio of aggregate solid mass to solid mass of intra-aggregate matrix (dimensionless)
$k$      lacunar factor entering Eqs.(5) and (6) (dimensionless)
$p$      porosity of silt and sand grains coming into contact (dimensionless)
$U$      specific volume of a soil intra-aggregate matrix per unit mass of the oven-dried matrix itself; $U(w)$ is the auxiliary shrinkage curve ($cm^3 g^{-1}$)
$U_a$      specific volume of soil aggregates per unit mass of oven-dried soil ($cm^3 g^{-1}$)
$U_{ah}$      $U_a(w_h')$ value ($cm^3 g^{-1}$)
$U_{az}$      oven-dried $U_a$ value ($cm^3 g^{-1}$)
$U_{cp}$      specific volume of clay pores in an intra-aggregate matrix per unit mass of the oven-dried matrix itself ($cm^3 g^{-1}$)
$U_h$      $U(w_h)$ value ($cm^3 g^{-1}$)
$U_i$      contribution of an interface aggregate layer to $U_a$ per unit mass of oven-dried soil ($cm^3 g^{-1}$)
$U_{lp}$      specific volume of lacunar pores in an intra-aggregate matrix per unit mass of the oven-dried matrix itself ($cm^3 g^{-1}$)
$U_{lph}$      $U_{lp}(w_h)$ value ($cm^3 g^{-1}$)
$U_s$      specific volume of structural pores per unit mass of oven-dried soil ($cm^3 g^{-1}$)
$U_z$      oven-dried $U$ value ($cm^3 g^{-1}$)
$U'$      contribution of an intra-aggregate matrix to $U_a$ and $Y$ per unit mass of oven-dried soil ($cm^3 g^{-1}$)
$U_h'$      $U'(w_h')$ value ($cm^3 g^{-1}$)
$U_z'$      oven-dried $U'$ value ($cm^3 g^{-1}$)
$V$      current value of the specific volume of a clay matrix ($cm^3 g^{-1}$)
$V_{cp}$      specific volume of clay matrix pores ($cm^3 g^{-1}$)
$V_z$      specific volume of a clay matrix in the oven-dried state ($cm^3 g^{-1}$)
$v_s$      relative volume of clay solids (dimensionless)

| | |
|---|---|
| $v_z$ | relative oven-dried clay volume (dimensionless) |
| $W$ | total gravimetric water content per unit mass of oven-dried soil (g g$^{-1}$) |
| $W_h$ | $W$ value at maximum aggregate (and soil) swelling or initial shrinkage point (g g$^{-1}$) |
| $W_h^*$ | $W$ value that corresponds to possible filling of lacunar pores (gg$^{-1}$) |
| $W_m$ | $W$ value that corresponds to possible filling of structural pores (g g$^{-1}$) |
| $W_m^*$ | $W$ value that corresponds to possible filling of lacunar and structural pores (gg$^{-1}$) |
| $W_n$ | $W$ value at the end point of the basic shrinkage area of a soil (g g$^{-1}$) |
| $W_s$ | $W$ value at the end point of the structural shrinkage area of a soil (g g$^{-1}$) |
| $W_z$ | $W$ value at the shrinkage limit of a soil (g g$^{-1}$) |
| $w$ | gravimetric water content of an intra-aggregate matrix per unit mass of oven-dried intra-aggregate matrix itself (g g$^{-1}$) |
| $w_h$ | $w$ at the maximum swelling point of the intra-aggregate matrix; $w_h = W_h$ (g g$^{-1}$) |
| $w_n$ | $w$ at the end point of the basic shrinkage of the intra-aggregate matrix (g g$^{-1}$) |
| $w_z$ | $w$ value at the shrinkage limit of the intra-aggregate matrix (g g$^{-1}$) |
| $w'$ | contribution of an intra-aggregate matrix to the total water content $W$ per unit mass of oven-dried soil (g g$^{-1}$) |
| $w_h'$ | $w'$ maximum at maximum aggregate swelling $W_h$; $w_h' < W_h$ (g g$^{-1}$) |
| $w_n'$ | $w'$ at the end point of the basic shrinkage of the soil, $W_n$; $w_n' = W_n$ (g g$^{-1}$) |
| $w_s'$ | $w'$ at the end point of the structural shrinkage of the soil, $W_s$; $w_s' = W_s$ (g g$^{-1}$) |
| $w_z'$ | $w'$ value at the shrinkage limit of the soil, $W_z$; $w_z' = W_z$ (g g$^{-1}$) |
| $\bar{w}$ | gravimetric water content of a clay matrix (g g$^{-1}$) |
| $\bar{w}_h, \bar{w}_M, \bar{w}_n, \bar{w}_z$ | $\bar{w}$ values at maximum swelling, liquid limit, air-entry point, and shrinkage limit, respectively (g g$^{-1}$) |
| $Y$ | specific volume of a soil (cm$^3$ g$^{-1}$) |
| $Y_z$ | oven-dried $Y$ value (cm$^3$ g$^{-1}$) |
| $\Pi$ | rigid clay porosity of an interface aggregate layer (dimensionless) |
| $\rho_s, \rho_w$ | density of solids and water (g cm$^{-3}$) |
| $\omega$ | contribution of interface layer to $W$ per unit mass of oven-dried soil (g g$^{-1}$) |
| $\omega_h$ | $\omega$ maximum at maximum aggregate swelling $W_h$ (g g$^{-1}$) |

**Figure captions**

**Fig.1**. Qualitative view of (a) the clay and (b) soil shrinkage curves (modified from Chertkov, 2007, Fig.1). Subscripts "z", "n", "s" and "h" correspond to the shrinkage limit, air-entry point, final point of structural shrinkage, and maximum swelling point, respectively. The curve slope in the basic shrinkage area for clay is always unity, but for soil it can also be less than unity. Curves 1 and 2 are the observed variants in the structural shrinkage area. Dashed and dash-dot lines are the true and pseudo saturation lines, respectively. $(W_h^*-W_h)/\rho_w$ gives the volume of non-connected isolated pores that remain non-water filled.

**Fig.2**. Illustrative scheme of the internal structure of aggregates at (a) a clay content $c>c^*$ and at (b) $c<c^*$ where $c^*$ is the critical soil clay content (reproduced from Chertkov, 2007, Fig.2).

**Fig.3**. Qualitative view of the shrinkage curve $U(w)$ of the intra-aggregate matrix (corresponding to the shrinkage curve $V(\overline{w})$ of clay contributing to the matrix in Fig.1a; with $w=\overline{w}c$). Lacunar factor $k$ is in the range $0 \leq k \leq 1$. The $k$ value depends on $c>c^*$ (Fig.2a) or $c<c^*$ (Fig.2b) and the lacunar pore volume in the last case. In the basic shrinkage area, $w_n<w<w_h$ the slope $dU/dw=(1-k)/\rho_w<1/\rho_w$. At $k>0$ the shrinkage curve, $U(w)$ starts at the pseudo saturation (dash-dot) line; the $(w_h^*-w_h)$ shear relative to the true saturation (dashed) line is connected with non-filled lacunar pores.

**Fig.4**. Transition from the shrinkage curve, $U(w)$ of the intra-aggregate matrix to the contribution, $U'(w')$ of the intra-aggregate matrix to the specific volume of aggregates ($U_a$) or soil ($Y$), for clay contents $c<c^*$, based on Eqs.(16)-(18). In particular, $w'_z=w_z/K$, $w'_n=w_n/K$, $w'_h=w_h/K$. Dashed and dash-dot lines are as in Fig.1.

**Fig.5**. Transition from the contribution, $U'(w')$ of the intra-aggregate matrix to the aggregate shrinkage curve in coordinates $U_a$ and $w'$ at $c<c^*$, based on Eq.(15) and accounting for the condition of Eq.(19). Dashed and dash-dot lines are as in Fig.1.

**Fig.6**. The qualitative view of two simplest possible variants of the $\omega(w')$ dependence (entering Eq.(14)). $\omega_h$ is the maximum water content of interface layer of aggregates. $w_s'$ corresponds to the completion of the structural shrinkage.

**Fig.7**. Transition from the aggregate shrinkage curve in coordinates $U_a$ and $w'$ to that in coordinates $U_a$ and $W$, at $c<c^*$. Dashed and dash-dot lines are as in Fig.1.

**Fig.8**. Transition from the aggregate shrinkage curve, $U_a(W)$ to the soil reference shrinkage curve, $Y(W)$, at $c<c^*$, based on Eq.(22). Dashed and dash-dot lines are as in Fig.1.

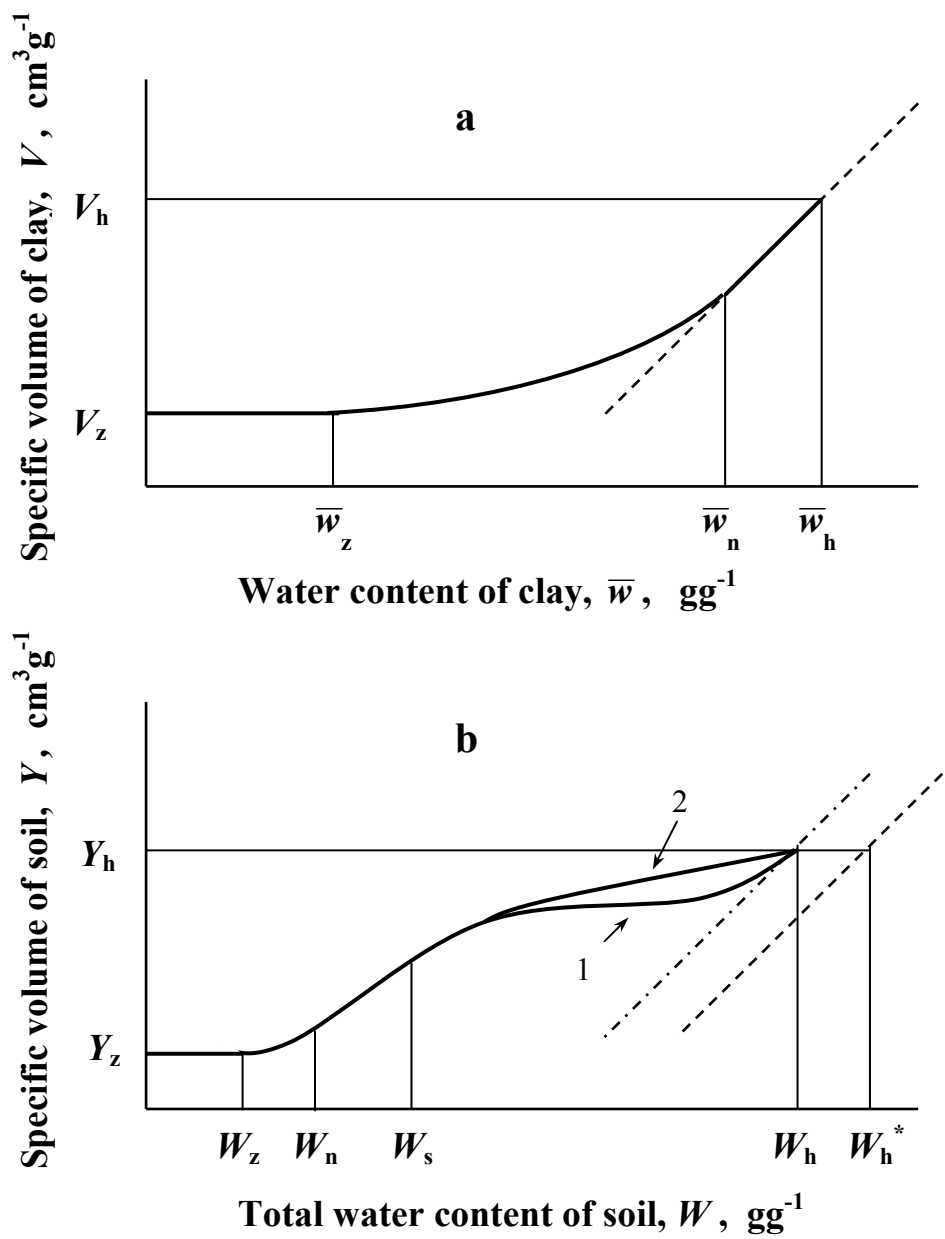

Fig.1

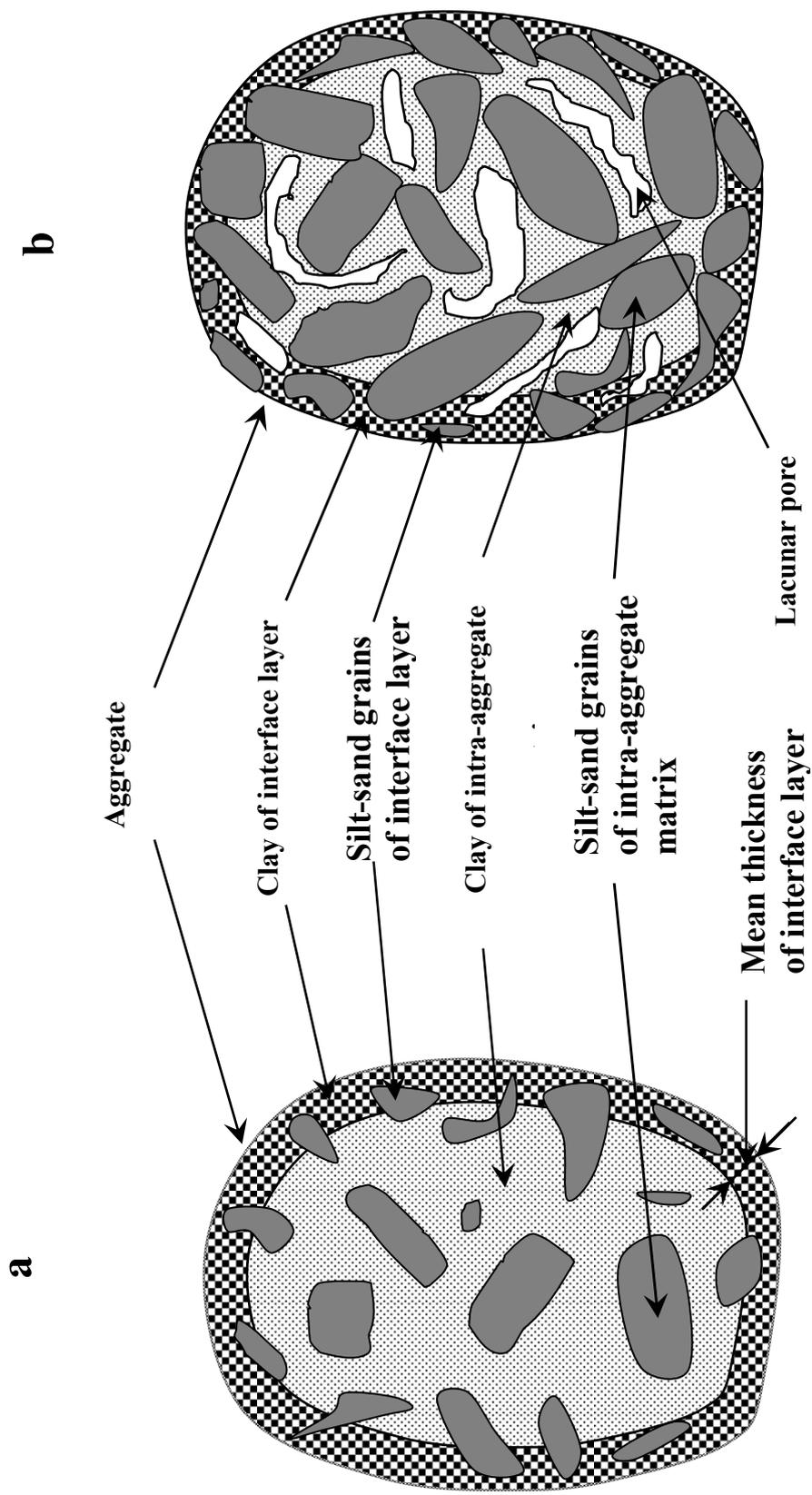

Fig.2

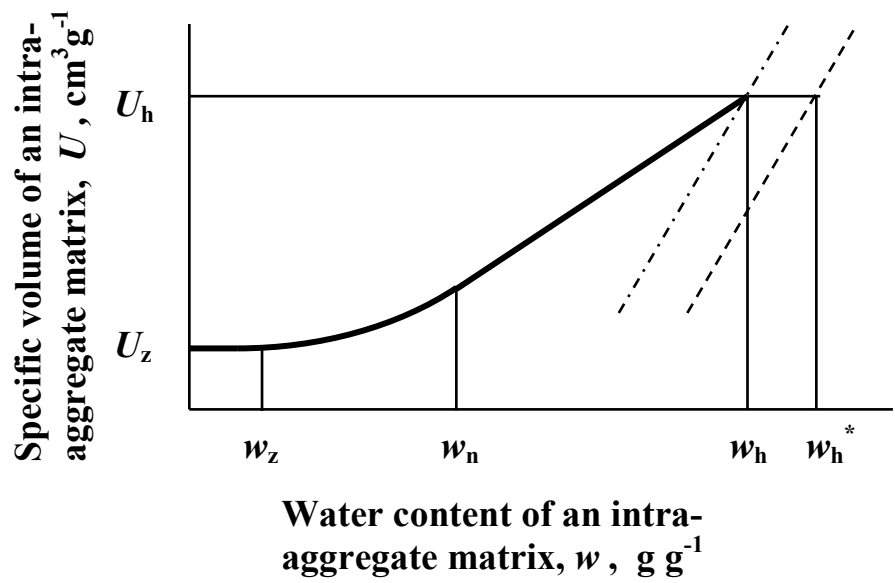

Fig.3

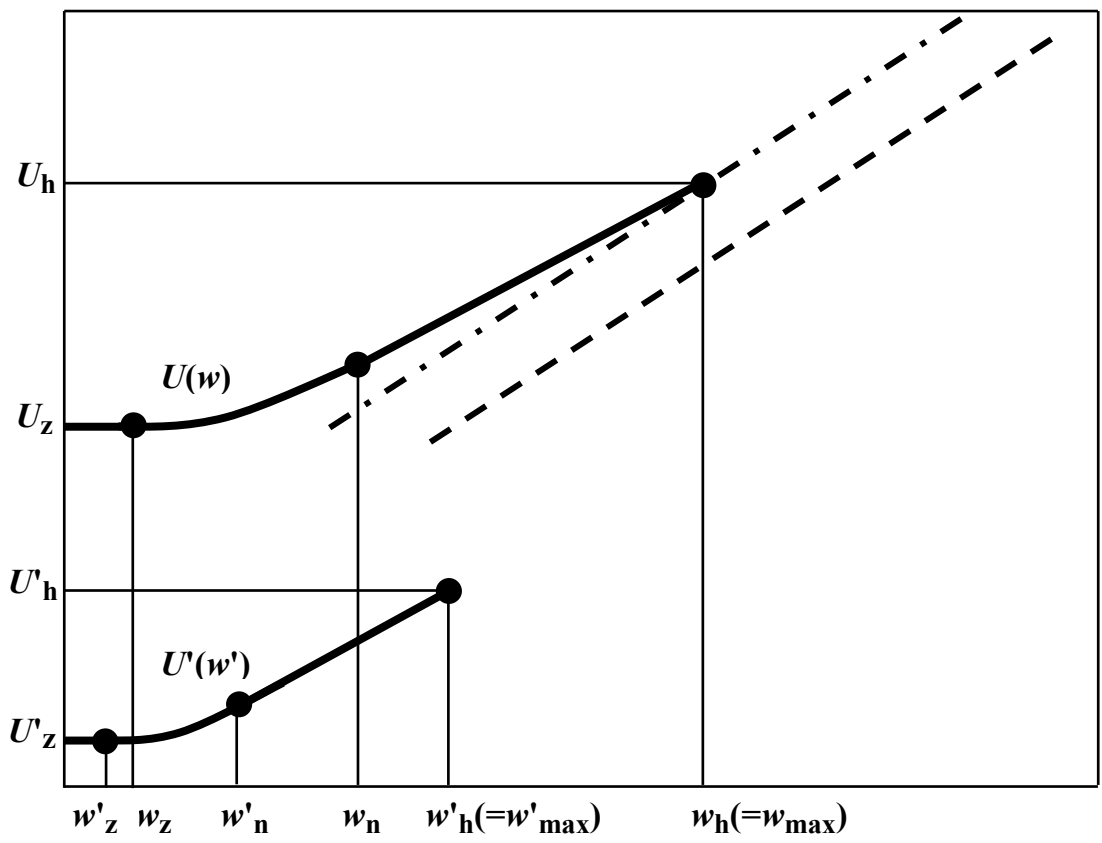

Fig.4

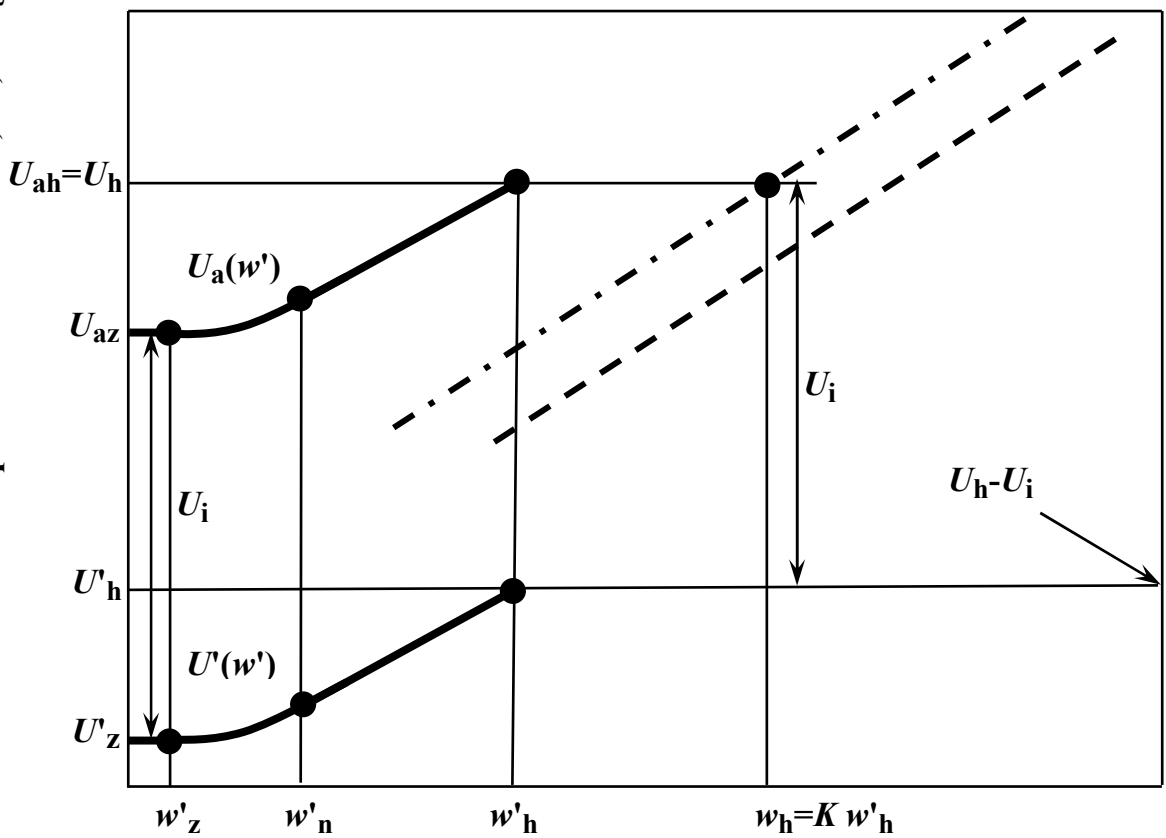

Fig.5

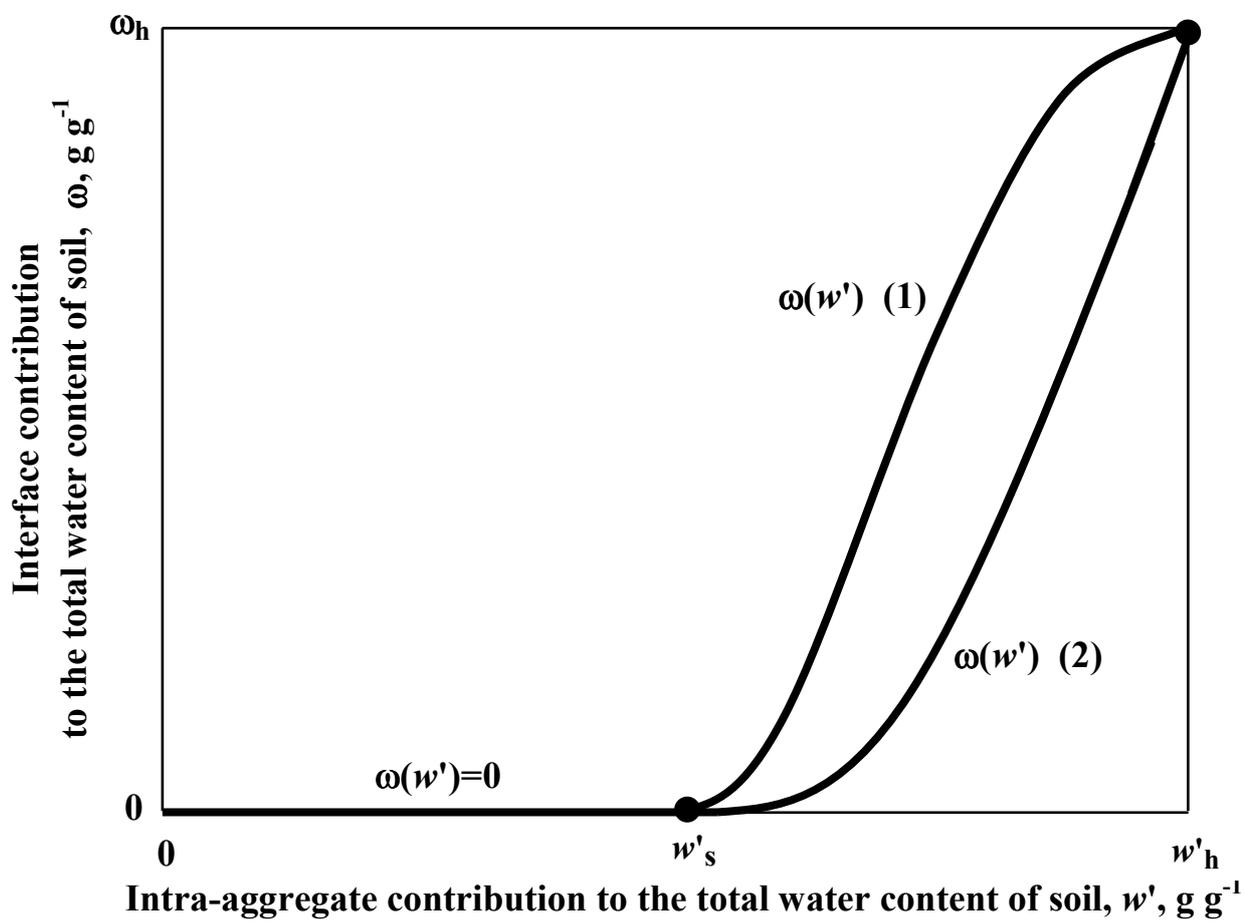

Fig.6

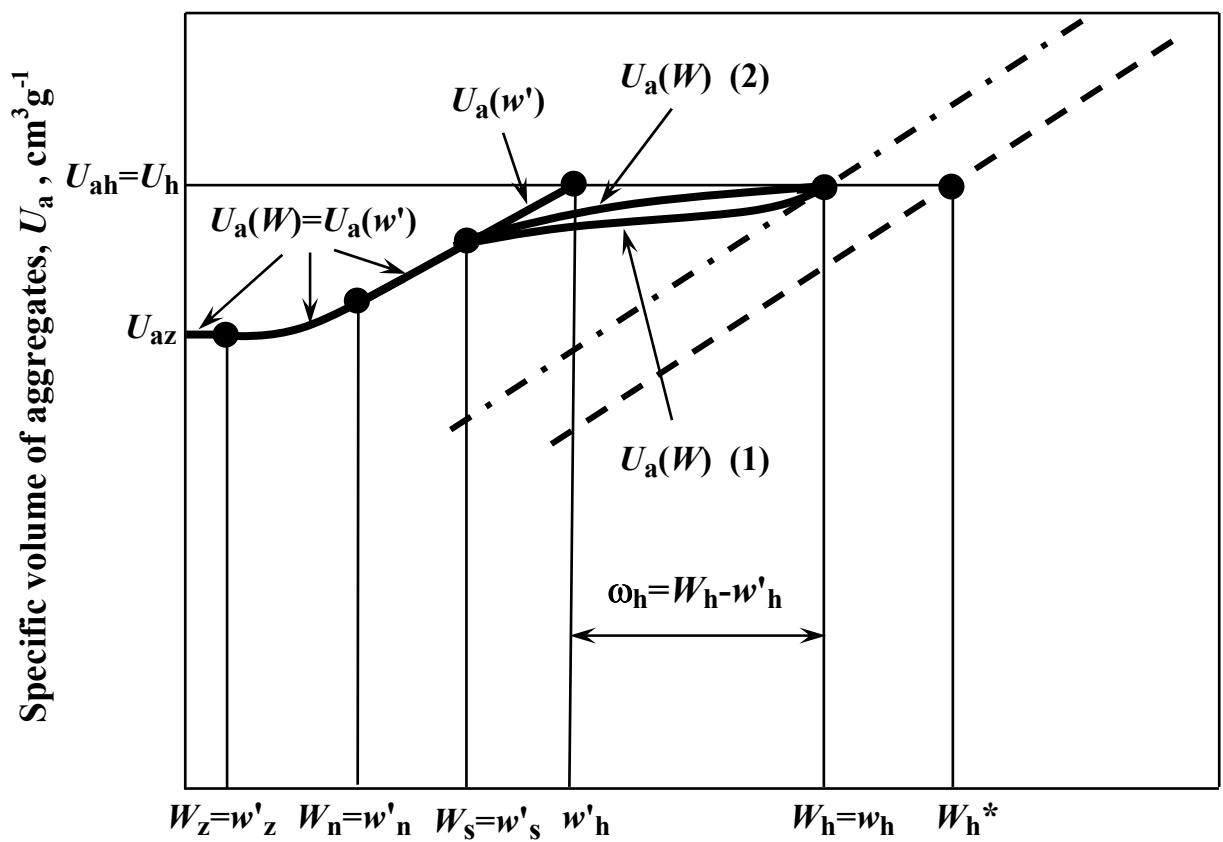

Fig.7

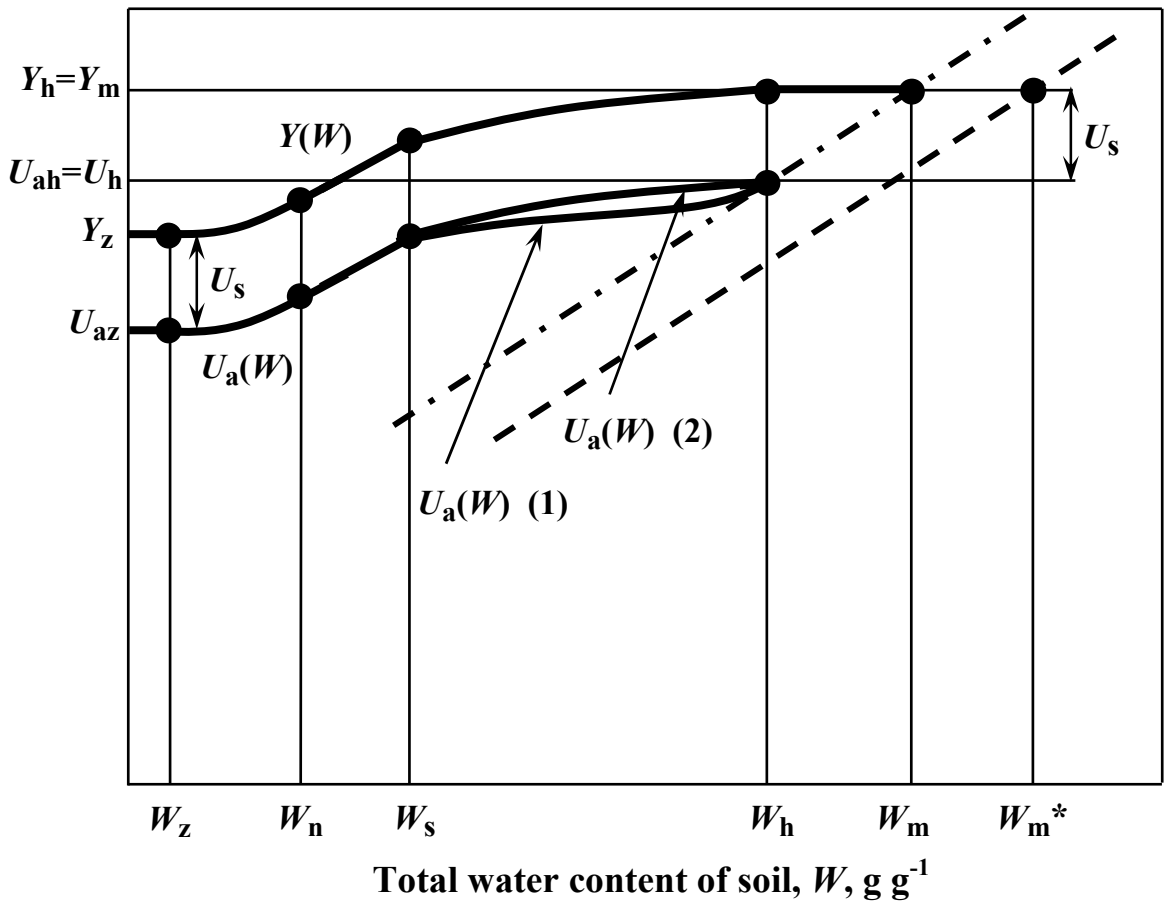

Fig.8